\documentclass[12pt]{article}
\usepackage{graphicx}
\usepackage{amssymb}
\usepackage{amsmath}
\numberwithin{equation}{section}

\newcommand{\eh}{\hfill}\newlength{\sperr}

\def\d{\delta}

\def\ve{\varepsilon}
\def\nn{\nonumber}

\newtheorem{Pa}{Paper}[section]
\newtheorem{Tm}[Pa]{{\bf Theorem}}

\newtheorem{Rk}[Pa]{{\bf Remark}}

\newtheorem{Dn}[Pa]{{\bf Definition}}

\title{Laws of thermodynamics and game theory}
\author{Lev Sakhnovich}
\date{}
\begin{document}
\maketitle

\thanks{99 Cove ave., Milford, CT, 06461, USA \\
 E-mail: lsakhnovich@gmail.com}\\
 
 
 dynamics, entropy,
 energy, residual entropy, ground state, game theory. \\

\begin{abstract} Using a game theory approach and 
a new extremal problem, Gibbs formula is proved  in a most simple and
general way for the classical mechanics case.  A corresponding conjecture
on the asymptotics of the classical entropy is formulated.
For the ordinary quantum mechanics case,
the third law of thermodynamics is derived.
Some results on the number of ground states
and  residual entropy are obtained rigorously.
\end{abstract}

 \section{Introduction}
 Let the eigenvalues $E_{n}$
 of the energy operator  be given. Consider the mean quantum energy
 \begin{equation}\label{1}
 E_q=\sum_{n}P_{n}E_{n} \end{equation}
 and the quantum entropy
 \begin{equation}\label{2}
 S_q=-\sum_{n}P_{n}\log P_{n},\end{equation}
 where $P_{n}$ are the corresponding probabilities, that is,
 \begin{equation}\label{2'} \sum_{n}P_{n}=1.\end{equation}
 The following statement and definition are well-known. 
 
 \textbf{Second Law of Thermodynamics} (see, e.g., \cite{St}).
 \emph{Let the eigenvalues $E_{n}$ and mean energy $E_q$ be fixed.
 Then the entropy $S_q$ takes the maximal value.}
 
 \begin{Dn}\label{Dn1}
$($see, e.g., \cite{My}$)$.
Game theory models strategic situations in which an individual's success in making choices depends  on the choices of
 others.
 \end{Dn}
 
 According to the second law of thermodynamics a problem of the \emph{conditional} extremum appears.
 Since the extremum is conditional, the connections between the mean energy $E_q$ and entropy $S_q$ can be interpreted in terms of the game theory. We introduce the "compromise" function
 \begin{equation}\label{3}
 F={\lambda}E_q+S_q,\quad \lambda=-\beta=-1/(kT),\end{equation}
 where k is Bolzmann constant, T is temperature, $\{E_{n}\}$ and $\lambda$ are fixed.
 
 {\bf Fundamental principle.} The function F defines the game between the mean energy
$E_q$ and entropy $S_q$.

In other words, the maximum point $P_{max}=(P_{1},P_{2},\ldots,P_{n},\ldots)$ of F gives the
corresponding probabilities and defines the compromise values of $E_q$ and $S_q$.
We  proved in  \cite{LAS3} that
\begin{equation}
\label{4}
P_{n}=e^{\lambda{E_n}}/Z_q, \quad Z_q=\sum_{n}e^{\lambda{E_n}}.
\end{equation}
Thus, we deduced the basic formulas  \eqref{1}, \eqref{2}, and \eqref{4} for the mean energy and entropy (quantum case) using the game theory approach. In the present note we apply the game theory approach to the classical case 
(Section \ref{SecClC}). Section  \ref{Sec3law}  is dedicated 
to the third law of thermodynamics (classical and quantum cases).
For that purpose, we use the results from Introduction and Section \ref{SecClC}.
\section{Classical thermodynamics \\ and game theory}
\label{SecClC}
 Let us consider the classical case with
the classical Hamiltonian
\begin{equation}\label{5}
H(p,q)=\frac{1}{2m}\sum_{j=1}^{N}p_{j}^{2}+V(q),\end{equation}
where $p=(p_{1},p_{2},...,p_{N})$, $\,q=(q_{1},q_{2},...,q_{N})$, $p_{j}$ are generalized
momenta, and $q_{j}$ are generalized coordinates. The classical mean energy 
$E_{c}$
and the classical entropy $S_{c}$ are defined by the relations
\begin{equation}\label{6}
E_{c}=\int\int{H(p,q)}{\it P}(p,q)dpdq,\end{equation}
\begin{equation}
\label{7}
S_{c}=-\int\int{\it P}(p,q)\log{\it P}(p,q)dpdq,\end{equation}
where
\begin{equation}\label{8}
{\it P}(p,q){\geq}0, \quad \int\int{\it P}(p,q)dpdq=1.\end{equation}
In the classical case we  consider again the game between energy $E_{c}$ and entropy $S_{c}$. The parameter $\beta=1/kT=-\lambda$ is fixed.
In a way, which is quite similar to \eqref{3},
we introduce the compromise function
\begin{equation}\label{9}
F_{c}={\lambda}E_{c}+S_{c},\end{equation}
where the parameter $\beta=1/kT=-\lambda$ is fixed.
Next, we use the calculus
of variations. The corresponding Euler's equation takes the form
\begin{equation}\label{10}
\frac{\d}{\d {\it P}}\big({\lambda}H(p,q){\it P}(p,q)-{\it P}(p,q)\log{\it P}(p,q)+
{\mu}{\it P}(p,q)\big)=0.\end{equation}
Here $\frac{\d}{\d {\it P}}$ stands for the functional derivation,
$\mu$ is the Lagrange multiplier, and our extremal problem is conditional
(see \eqref{8}). Because of \eqref{10} we have
\begin{equation}\label{11}
{\lambda}H(p,q)-1- \log{\it P}(p,q)+
{\mu}=0.\end{equation}
From \eqref{11} we obtain
\begin{equation}
\label{12}{\it P}(p,q)=Ce^{{\lambda}H(p,q)}.\end{equation}
Formulas \eqref{8} and \eqref{12}
imply that
\begin{equation}\label{13}
{\it P}(p,q)=e^{{\lambda}H(p,q)}/Z_{c}, \quad
Z_{c}=\int\int e^{\lambda{H(p,q)}}dpdq
\end{equation}
\begin{Rk}\label{Rk2.1}
 The famous formula \eqref{13} was deduced above.
 We think that it was done in the simplest way. Note that
\begin{equation}\label{14}
\frac{\d^2}{\d {\it P}^2}F_{c}=-1/{\it P}<0.\end{equation}
It means that, under condition \eqref{8}, the functional $F_c$ of the form    
\eqref{9} attains maximum
for ${\it P}(p,q)$, which is defined by formula \eqref{13}.
\end{Rk}
\section{Third law of thermodynamics}
\label{Sec3law}
\subsection{Quantum case}
We suppose that $h$ in the energy operator is fixed
and its eigenvalues $E_n=E_n(h)$ are indexed so that
\begin{equation}\label{3.1}
E_{1} \, {\leq}\, E_{2} \, {\leq} \, E_{3}\, \leq \ldots \, .
\end{equation}
Similar to \eqref{4} we assume that the statistical sum $Z_q$ is bounded:
\begin{equation}\label{3.2}
Z_{q}(\beta)=\sum_{n=1}^{\infty}e^{-\beta{E_{n}}}<\infty. \end{equation}
For simplicity, we assume that \eqref{3.2} holds for all $\beta >0$.
Since for every $\ve >0$ there is an $N_{\ve}$, such that
\begin{equation}\label{r1}
0<E_n e^{-\ve E_{n}}<1 \quad {\mathrm{for \,\, all}} \quad n>N_{\ve},
\end{equation}
inequality \eqref{3.2} for all $\beta >0$ implies
\begin{equation}\label{r2}
\sum_{n=1}^{\infty}E_n e^{-\beta{E_{n}}}<\infty \quad
 {\mathrm{for \,\, all}} \quad \beta >0.
\end{equation}
Therefore, formulas \eqref{1}, \eqref{3}, and \eqref{4} lead us to 
the equality
\begin{equation}\label{3.3}
E_{q}(\beta)=\sum_{n=1}^{\infty}E_{n}e^{-\beta{E_{n}}}/Z_{q}(\beta). \end{equation}
From \eqref{3.2},   \eqref{r2}, and \eqref{3.3} we deduce the following relations:
\begin{align}
\nn & \sum_{n=1}^{\infty}E_{n}e^{-\beta{E_{n}}}=e^{-\beta E_{1}}(mE_1+O(e^{-\beta(E_{m+1}-E_{1})})),
\quad \beta{\to}\infty,\quad \beta=1/kT,
\\ \label{3.5}&
Z_{q}(\beta)=e^{-\beta E_{1}}(m+O(e^{-\beta(E_{m+1}-E_{1})})),\quad \beta{\to}\infty,\quad \beta=1/kT,
\\ \label{3.4}&
E_{q}(\beta)=E_{1}+O(e^{-\beta(E_{m+1}-E_{1})}), \quad \beta{\to}\infty,\quad \beta=1/kT,
\end{align}
where $m$ is the multiplicity of $E_{1}$.

Equalities \eqref{2}, \eqref{3}, and \eqref{4} imply a formula for entropy
\begin{equation}\label{3.6}
S_{q}(\beta)={\beta}E_{q}(\beta)+\log Z_{q}(\beta).
\end{equation}
Using relations \eqref{3.5}-\eqref{3.6} we obtain 
\begin{equation}\label{3.7}
S_{q}(\beta){\to}\log (m),\quad \beta{\to}\infty,\quad \beta=1/kT.\end{equation}
Compare \eqref{3.7} with the well-known statement:

{\bf Third law of thermodynamics.} 
If  $\beta{\to}\infty$, then $S_{q}(\beta){\to}0$.

Thus, we proved the following assertion.
\begin{Tm}\label{Tm1}
Let the conditions \eqref{3.1} and \eqref{3.2} be fulfilled.
Then the relation $m=1$ and the
third law of thermodynamics are equivalent.
\end{Tm}
\begin{Rk}
The equality $m=1$ means that the ground  state is non-degenerate.
\end{Rk}
\begin{Rk}
 If $m>1$ we obtain the so called residual entropy $\log (m)$ $($see \cite{KL}$)$.
\end{Rk}
\subsection{Classical case}
Now, we consider briefly the third law of   thermodynamics for the classical case. 
First, assume that the dimension $N$ of the coordinate space equals 1.
In case of a potential well the following formulas:
\begin{equation}\label{3.8}
E_{c}(\beta)=1/(2\beta),\quad Z_{c}(\beta)=\sqrt{2{\pi}ma^{2}/\beta}
\end{equation} 
hold (see \cite{LAS1, LAS2}). The corresponding formulas for the oscillator 
(see \cite{LAS1}) have the form
\begin{equation}\label{3.9}
E_{c}(\beta)=1/\beta,\quad Z_{c}(\beta)=2{\pi}/(\beta\omega).
\end{equation}
It follows from \eqref{6},  \eqref{7}, and \eqref{13} that
\begin{equation}\label{3.10}
S_{c}(\beta)={\beta}E_{c}(\beta)+\log Z_{c}(\beta) \end{equation}
Because of \eqref{3.10}, in both cases \eqref{3.8} and \eqref{3.9} we have
\begin{equation}\label{3.11}
S_{c}(\beta)=c_{1}+c_{2}\log {\beta},\end{equation}
where $c_{1}$ and $c_{2}$ are constants. Note that relation \eqref{3.11} holds 
also for an arbitrary $N$ (the corresponding formulas for the potential well
are adduced in \cite{LAS3}).
In view of \eqref{3.11}, we formulate our conjecture:

\textbf{Conjecture 1.} In the  classical case the following result
\begin{equation}\nonumber
S_{c}(\beta)=c_{1}+c_{2}\log {\beta}+o(1),\quad \beta{\to}\infty
\end{equation}
is valid.

\end{document}